\documentclass[twocolumn, twocolappendix]{aastex631}

\newcommand{\cii}{[CII]}
\newcommand{\lcii}{$L_{\rm [CII]}$}

\newcommand{\kms}{km s$^{-1}$}
\newcommand{\lbol}{$L_{\rm Bol}$}
\newcommand{\lsun}{$\rm L_{\odot}$}
\newcommand{\msun}{$\rm M_{\odot}$}
\newcommand{\psoj}{PSOJ183$+$05}
\newcommand{\ergs}{erg s$^{-1}$}
\newcommand{\mjybeam}{mJy beam$^{-1}$}
\newcommand{\mujybeam}{$\mu$Jy beam$^{-1}$}
\newcommand{\jybeamkms}{Jy beam$^{-1}$ km s$^{-1}$}

\shorttitle{CGM emission and a biconical outflow in a $z\sim6.4$ quasar}
\shortauthors{Bischetti et al.}
\graphicspath{{./}{images/}}
\usepackage{CJKutf8}
\usepackage{multirow}
\usepackage{float}

\begin{document}

\begin{CJK}{UTF8}{}
\CJKfamily{mj}

\title{ALMA reveals bright circumgalactic emission and a biconical outflow in $\mathbf{z\sim6.4}$ quasar PSOJ183$+$05}

\correspondingauthor{M. Bischetti}
\email{manuela.bischetti@units.it}

\author[0000-0002-4314-021X]{Manuela Bischetti}
\affiliation{Dipartimento di Fisica, Universit\'a di Trieste, Sezione di Astronomia, 
Via G.B. Tiepolo 11, I-34131 Trieste, Italy}
\affiliation{INAF - Osservatorio Astronomico di Trieste, Via G. B. Tiepolo 11, I–34131 Trieste, Italy}

\author[0000-0002-4227-6035]{Chiara Feruglio}\affiliation{INAF - Osservatorio Astronomico di Trieste, Via G. B. Tiepolo 11, I–34131 Trieste, Italy}
\affiliation{IFPU - Institut for fundamental physics of the Universe, Via Beirut 2, 34014 Trieste, Italy }

\author[0000-0002-6719-380X]{Stefano Carniani}\affiliation{Scuola Normale Superiore, Piazza dei Cavalieri 7, I-56126 Pisa, Italy}

\author[0000-0003-3693-3091]{Valentina D'Odorico}
\affiliation{INAF - Osservatorio Astronomico di Trieste, Via G. B. Tiepolo 11, I–34131 Trieste, Italy}
\affiliation{IFPU - Institut for fundamental physics of the Universe, Via Beirut 2, 34014 Trieste, Italy }
\affiliation{Scuola Normale Superiore, Piazza dei Cavalieri 7, I-56126 Pisa, Italy}

\author[0000-0003-4751-7421]{Francesco Salvestrini}\affiliation{INAF - Osservatorio Astronomico di Trieste, Via G. B. Tiepolo 11, I–34131 Trieste, Italy}

\author[0000-0002-4031-4157]{Fabrizio Fiore}\affiliation{INAF - Osservatorio Astronomico di Trieste, Via G. B. Tiepolo 11, I–34131 Trieste, Italy}
\affiliation{IFPU - Institut for fundamental physics of the Universe, Via Beirut 2, 34014 Trieste, Italy }

\begin{abstract}
Understanding gas flows between galaxies and their surrounding circum-galactic medium (CGM) is crucial to unveil the mechanisms regulating galaxy evolution, especially in the early Universe. However, observations of the CGM around massive galaxies at $z>6$ remain limited, particularly in the cold gas phase. In this work, we present multi-configuration ALMA observations of [CII]$\lambda158,\mu$m and millimetre continuum emission in the $z\sim6.4$ quasar \psoj. We find clumpy [CII] emission, tracing gas up to a $\sim6$ kpc radius, consistent with the interface region between the interstellar medium (ISM) and CGM. The [CII] kinematics shows a rotating disk and a high-velocity, biconical outflow extending up to 5 kpc. The inferred  mass outflow rate is $\dot{M}_{\rm of}\sim930$ M$\odot$ yr$^{-1}$, among the highest at $z>6$, and comparable to the star-formation rate. These findings suggest that quasar-driven outflows can rapidly transfer energy and momentum to the CGM, without immediately quenching star formation in the host galaxy ISM. This supports a delayed feedback scenario, in which outflows reshape CGM conditions and regulate future gas accretion over longer timescales. We find that neither the high-velocity component nor the extended CGM emission in \psoj\ are recovered when using the high-resolution dataset alone, which may explain the conflicting results reported regarding \cii\ sizes and the detection of outflows at $z\gtrsim6$.
Combining multi-configuration ALMA data with observations from JWST and MUSE will be crucial to map the CGM across its different phases and build a comprehensive picture of the baryon cycle in the first massive galaxies.

\end{abstract}



\section{Introduction} \label{sec:intro}

The acquisition, ejection, and recycling of gas are fundamental processes driving galaxy evolution. The primary site of these gas flows is the circum-galactic medium (CGM), a region extending beyond a galaxy’s stellar distribution up to its virial radius and beyond. CGM serves as a major baryon reservoir, providing inflows of fuel for star-formation to the insterstellar medium (ISM), and acting as the immediate destination for outflows driven by feedback processes within galaxies \citep{Tumlinson17}. Attaining a complete picture of how galaxies evolve is thus hampered by our incomplete knowledge of the cycle of baryons between ISM and CGM.  
This issue is particularly relevant for the first massive ($\gtrsim10^{10}$ \msun) galaxies, which formed 0.5-1 Gyr after the Big Bang ($z>6$). According to the hierarchical scenario of structure formation \citep[e.g.][]{Springel05}, these galaxies trace the densest regions of the Universe which were the first to collapse, and thus represent a relevant population of progenitors of the most massive galaxies that we observe in today’s Universe. 

The first pioneering studies of CGM halos relied on absorption features of the intervening gas in quasar spectra, and provided a robust description of the average CGM conditions from the first Gyr \citep[e.g.][]{Adelberger05, Turner14, Prochaska14, Kashino23, Galbiati23}. 
However, these probes precluded direct constraints on the three-dimensional structure of the CGM. Spatially resolved observations of warm ($10^4-10^5$ K) CGM emitting gas have been performed with sensitive integral field unit (IFU) spectroscopy using the Multi Unit Spectroscopic Explorer (MUSE). Indeed, CGM halos have been seen routinely glowing in Ly$\alpha$ around massive galaxies hosting quasars up to $z>6$ \citep{Borisova16,Farina19}. However, millimetre studies (mostly with the Atacama Large Millimetre Array, ALMA) have so far provided limited statistics on the detection of cold ($\lesssim10^2$ K) CGM halos  around a few massive galaxies and quasars at $z\sim2$ \citep{Cicone21, Jones23, Scholtz23} and at $z>6$ \citep{Cicone14, Lambert23, Meyer22, Bischetti24}. Studies based-on stacking have also provided conflicting results \citep{Fujimoto19, Novak20}. This scarcity of detections is likely due to most studies relying on observations not optimized to detect the diffuse CGM gas \citep[e.g. see discussion in][]{Carniani20}.

Different scenarios have been proposed to explain the origin of CGM halos around the first massive galaxies, including stellar and black-hole driven outflows able to push metal-enriched gas to large scales \citep{Costa22,Pizzati23}. IFU spectroscopy with the James Webb Space Telescope (JWST) has opened a new window into detecting and mapping warm outflows using the [OIII]$\lambda5008$ \AA\ emission line at $z>6$ \citep[e.g.][]{Marshall23}. However, probing outflows in the cold gas phase immediately preceding star formation remains an observational challenge, and only a few detections have been reported \citep{Maiolino12, Feruglio17, Bischetti19, Izumi21, Izumi21a, Tripodi22}. This is likely due to the limited sensitivity of most millimetre observations, considering similar outflowing to total gas ratios as observed in the low redshift Universe \citep{Bischetti19pds, Fluetsch19}.

In this work, we present ALMA observations of the [CII]$\lambda158$ $\mu$m emission line and millimetre continuum emission in the host galaxy of quasar \psoj\ at $z\sim6.4$ \citep{Banados16} and CGM environment. We combined archival observations acquired with different antenna configurations, to boost the detectability of diffuse cold gas with respect to previous studies and, at the same time, map it with high angular resolution. The target of this work is a bright quasar with a bolometric luminosity log($L_{\rm Bol}/\rm erg\ s^{-1}$)$\simeq47.2$, powered by a black hole with mass log($\rm M_{\rm BH}/M_\odot$)$\simeq9.4$ and hosting a nuclear wind, as traced by a large blueshift of the CIV$\lambda1549$ \AA\ emission line \citep{Farina22,Mazzucchelli23}. A molecular outflow arising from the central kpc region has been detected in OH$\lambda119$ $\mu$m absorption by \cite{Butler23}, with a median velocity of 530 km/s and an outflow rate of about 75-800 M$_\odot$ yr$^{-1}$. Previous \cii\ studies of this source reported a bright emission, with a high luminosity \lcii$\simeq7\times10^{9}$ \lsun\ \citep{Decarli18} associated with a relatively compact disk dominated by rotation \citep{Venemans20, Neeleman21}.
Throughout the paper, we adopt a spatially-flat ΛCDM cosmology with H$_0$ = 67.4 \kms, and $\Omega_M = 0.315$ \citep{Planck20}.






\section{ALMA Observations and Data analysis}\label{sect:data}

We analyse archival ALMA observations of \psoj\ targeting [CII] and band 6 continuum emission using CASA 6.4.0 software \citep{McMullin07}. We consider observations of the [CII] $\nu_{\rm rest}=1900.537$ GHz emission line, as acquired with  three different antenna configurations, including a low angular resolution (project ID 2015.1.01115.S, $\sim1.1$ arcsec), a high angular resolution (ID 2019.1.01633.S, $\sim0.12$ arcsec), and an intermediate resolution (ID 2016.1.00544.S, $\sim0.3$ arcsec) dataset. 

Combining visibilities acquired from different antenna configurations allows us to maximise the sensitivity to possible extended [CII] emission, while keeping a reasonably good angular resolution. We did not combine an additional low-resolution dataset (ID 2021.1.01082.S, $\sim1.3$ arcsec) available on the archive as its limited spectral coverage (velocities $|v|\lesssim1000$ \kms) with respect to the [CII] line would introduce high uncertainty in the modelling and subtraction of the continuum emission.

Visibilities were calibrated using the standard calibration provided by the ALMA observatory and the default phase, bandpass and flux calibrators. We merged the visibilities from the three datasets using CASA (version 6.6.5) task {\it concat}.  We created a  continuum map (Figure 1a) by averaging visibilities over all spectral windows, covering the observed frequencies $237.7-242.8$ and $252.7-257.9$ GHz, and excluding the spectral range covered by [CII]. To model and subtract the continuum emission from the line, we combined the adjacent spectral windows in the baseband containing [CII] and performed a fit in the $uv$ plane to channels with $|v|>1000$ \kms, using a first-order polynomial continuum. A continuum-subtracted datacube was created using CASA task $tclean$, with the $hogbom$ cleaning algorithm in non-interactive mode, a threshold equal to two times the rms sensitivity and a natural weighting of the visibilities. We adopted a 30 \kms\ channel width. The resulting synthesized beam for the spectral window including [CII] is 0.13$\times$0.12 arcsec$^2$, the rms sensitivity of the [CII] datacube is $\sigma_{\rm 30}=0.095$ \mjybeam\ for a 30 \kms\ channel width, and the rms sensitivity of the continuum map is $\sigma_{\rm cont}=6.6$ \mujybeam. 

\begin{table}[thb]
    \setlength{\tabcolsep}{2pt}
    \caption{Properties of \psoj.}
    \label{tab:cii-properties}
    \centering
    \begin{tabular}{lc}
    
    \hline\hline
    
    \hline
    RA & 12:12:26.974\\
    Dec & $+$05:05:33.540\\
    $z_{\rm [CII]}$ & 6.4388$\pm$0.0004 \\
    $S_{\rm cont}$ [mJy] & 3.53$\pm$0.04 \\
    $S_{\rm [CII]}$ [Jy \kms]    & 10.10$\pm$0.25\\
    L$_{\rm [CII]}$ [10$^{10}$L$_\odot$] & 1.09$\pm$0.03 \\
    $M_{\rm atom}$ [10$^{10}$M$_\odot$] & 1.51$\pm$0.04 \\

    \hline
    \multicolumn{2}{c}{[CII] outflow}\\
    $S_{\rm [CII]}$ [Jy \kms]    & 3.64$\pm$0.52\\    
    L'$_{\rm [CII]}$ [10$^{10}$L$_\odot$] & 0.39$\pm$0.05 \\
    $M_{\rm atom}$ [10$^{10}$M$_\odot$] & 0.54$\pm$0.04 \\
    $\langle v_{\rm mom1}\rangle$ [\kms] & $-30\pm65$ \\
    $\langle v_{\rm max}\rangle$ [\kms] & $790\pm110$ \\
    $r_{\rm of}$   [kpc] & $0.3-4.8$\\
    $\dot{M}_{\rm of}$ $[\rm M_\odot\ yr^{-1}]$ & $930^{+330}_{-290}$ \\

    \hline
    \end{tabular}
    
    \tablecomments{RA, Dec refer to the peak of the 237.7-257.9 GHz continuum. $\langle v_{\rm mom1}\rangle$ and $\langle v_{\rm max}\rangle$ are the flux weighted velocity shift and maximum velocity of the outflow, respectively (Fig. \ref{fig:outflow-maps}). Uncertainties correspond to a 68\% confidence level except for $\dot{M}_{\rm of}$ (see Sect. \ref{sect:outflow}).}
    
\end{table}

\begin{figure}[htb]
    \centering
    \includegraphics[width=1\columnwidth]{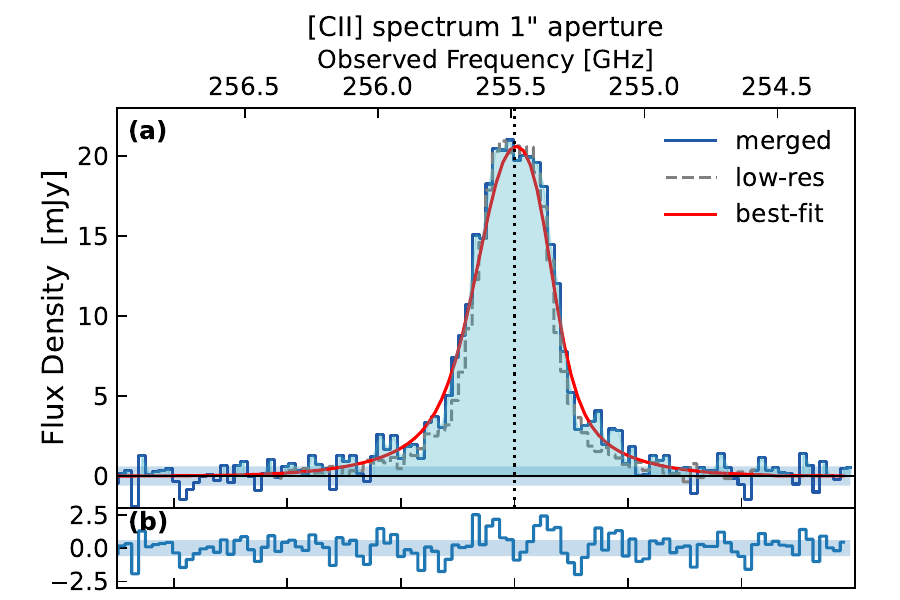}
    \includegraphics[width=1\columnwidth]{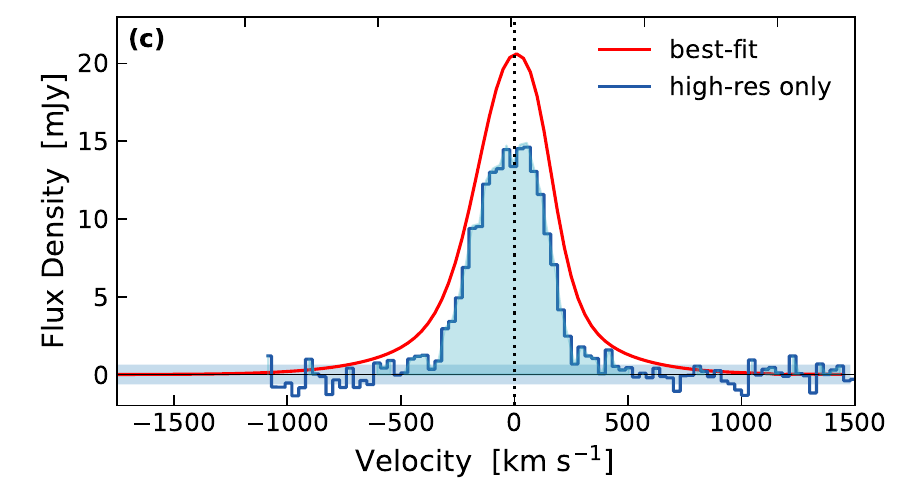}

    \caption{ (a): [CII] spectrum extracted from a circular 1 arcsec aperture (blue histogram), similar to the spatial extent of the broad [CII] emission component mapped in Fig. \ref{fig:outflow-maps}. The presence of blue/redshifted wings reaching velocities of $\pm1000$ \kms\ can be observed. These wings are also present in the spectrum extracted from the same aperture in the low-resolution dataset 2021.1.01082.S (normalized for comparison to the peak of the spectrum of the combined dataset). The best-fit profile, shown in red, is obtained through a pixel-by-pixel spectral decomposition using a two-Gaussian component model (Sect. \ref{sect:data}). (b): Residual spectrum obtained by subtracting the best-fit model to the data.
    The horizontal shaded region corresponds to the $\pm1\sigma$ noise level, calculated as  $\sigma_{\rm 30}\sqrt N$, where N is the number of the independent ALMA beams in the extraction aperture. (c) Spectrum extracted from the high-resolution dataset alone (Sect. \ref{sect:data}), showing no significant [CII] emission beyond $\pm500$ \kms.}
    \label{fig:cii-spectrum}    
\end{figure}

We performed a pixel-by-pixel spectral decomposition of [CII] emission detected at $>3\sigma_{\rm 30}$ in the continuum-subtracted datacube. 
To reproduce the [CII] emission line profile in \psoj\ (Fig. \ref{fig:cii-spectrum}a), we considered a model with two Gaussian components: a main one with FWHM$<500$ \kms\ to account for the systemic [CII] emission, based on previous line width measurements \citep{Venemans20}, and a second broad Gaussian with a FWHM$>500$ \kms, to account for possible high-velocity wings in the line profile. The amplitude of this broad component is limited at maximum 20\% of the systemic component, consistently with the most prominent [CII] wings observed in low-z active galaxies and high-z quasars \citep{Maiolino12,Bischetti19,Fluetsch19}. 

\begin{figure*}[thb]
    \centering
    \includegraphics[width = 0.49\linewidth]{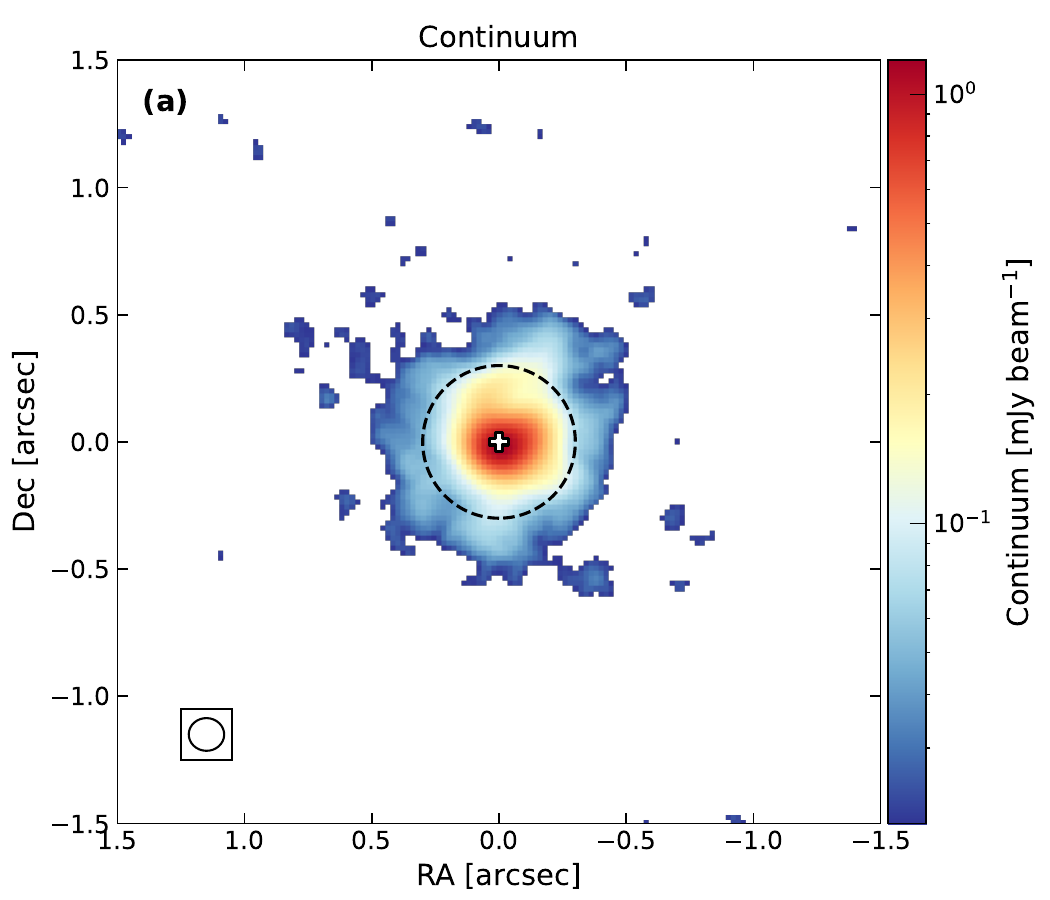}
    \includegraphics[width = 0.49\linewidth]{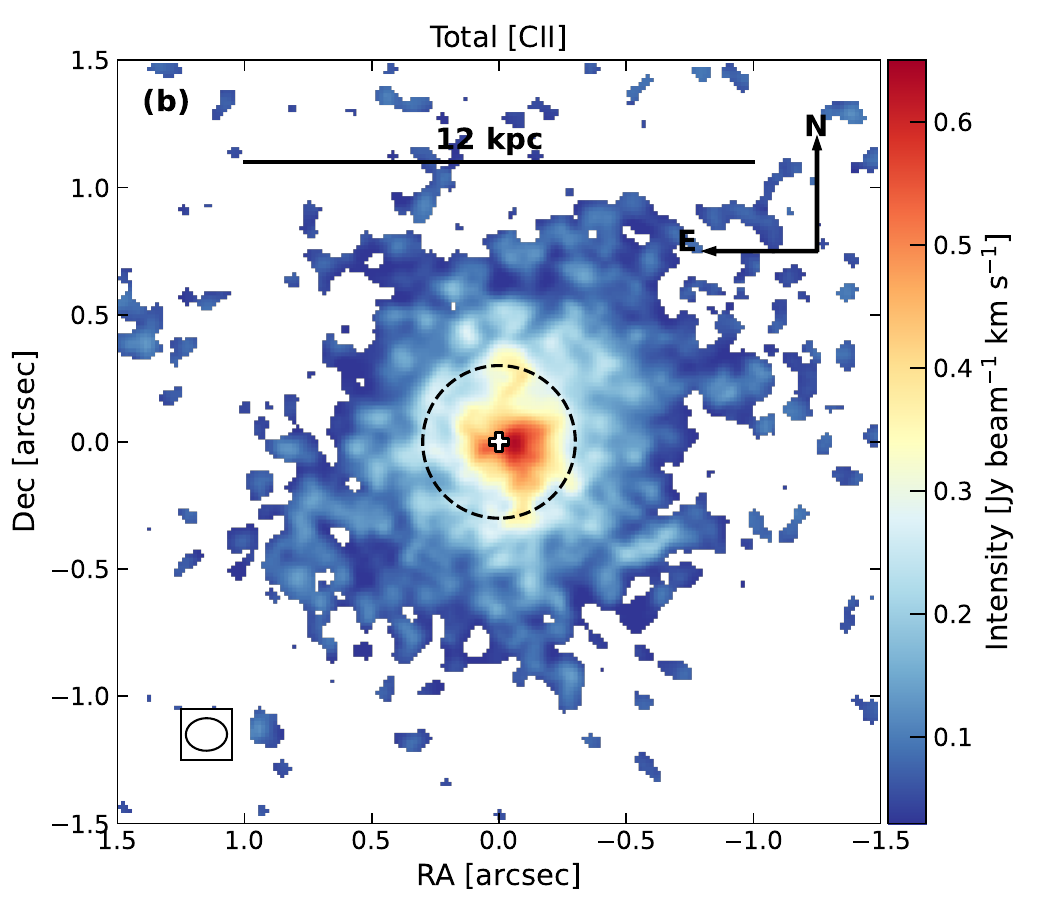}
    \includegraphics[width = 0.49\linewidth]{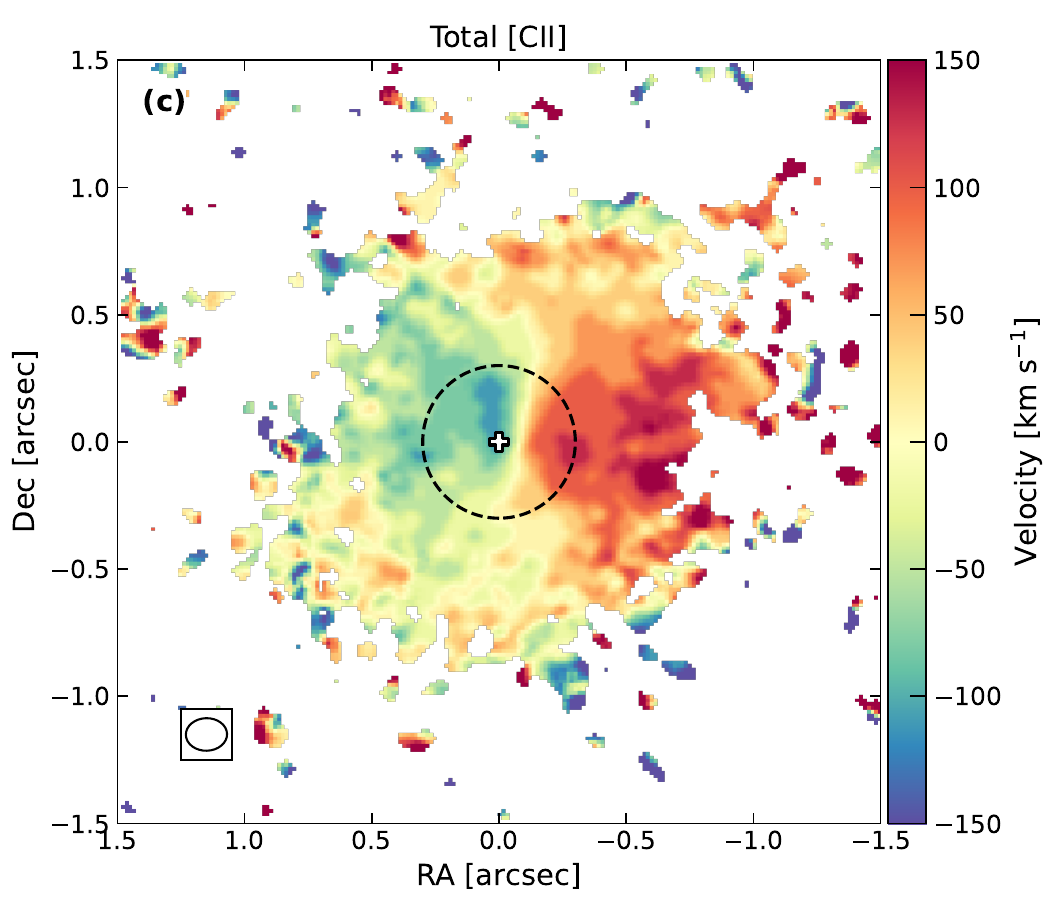}    
    \includegraphics[width = 0.49\linewidth]{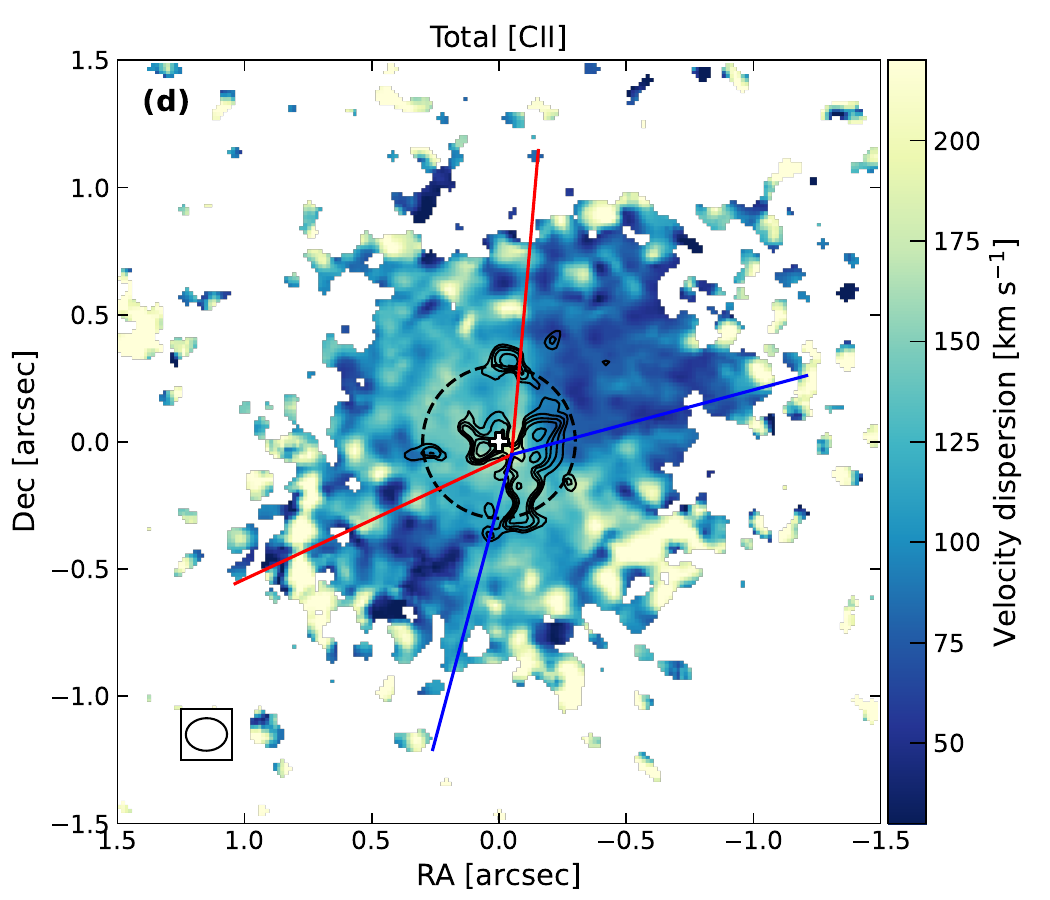}
    \caption{(a): Map of continuum emission detected at $>3\sigma_{\rm cont}$ in the host galaxy of \psoj. Quasar location, identified as the peak of the ALMA continuum, is shown by the cross. (b): Velocity-integrated intensity map of [CII] emission detected $>3\sigma_{\rm [CII]}$. (c): [CII] Velocity map.  (d) [CII] velocity dispersion map, showing a biconical region oriented along the NE-SW direction with high $\sigma_{\rm vel}\sim150$ \kms, highlighted by the overlaid double cone. Contours highlight the location of the high-velocity [CII] emission with intensity $>0.02$ Jy beam$^{-1}$ \kms\ (Fig. 3a).
    The displayed region of $3\times3$ arcsec$^2$ corresponds to the field of view covered by JWST/NIRSpec IFU. 
    The dashed circle indicates the size of the stellar distribution measured by JWST/NIRCam observations in the host-galaxies of $z\sim6$ quasars (Sect. \ref{sect:halo}). The white ellipse shows the ALMA beam.
}
    \label{fig:maps}
\end{figure*}

The resulting velocity-integrated intensity ($0^{\text{th}}$ moment), velocity ($1^{\text{st}}$ moment) and velocity dispersion ($2^{\text{nd}}$ moment) maps associated with the total [CII] emission in \psoj\ are shown in Fig. \ref{fig:maps}. Moment maps associated with the high-velocity [CII] emission are displayed in Fig. \ref{fig:outflow-maps}.

\section{Continuum and [CII] EMISSION}\label{sect:results}

ALMA observations of the $\sim240-255$ GHz continuum in \psoj, detected at $\sim160\sigma_{\rm cont}$ level, reveal that the bulk of the emission is associated with dust in the host galaxy ISM (Fig. \ref{fig:maps}a). By fitting the continuum map with a 2D Gaussian profile, we infer a deconvolved size of ($0.23\pm0.01$) $\times$ ($0.21\pm0.01$) arcsec$^2$, corresponding to about $1.3\times1.2$ kpc$^2$. This compact size is consistent with that previously measured for \psoj\ by \cite{Venemans20} using $0.2$ arcsec ALMA observations, and similar to those reported for the millimeter continuum in the host galaxies of most $z\gtrsim6$ quasars \citep[e.g.][]{Feruglio18,Tripodi24,Salvestrini25}. It is also consistent with the region in which the bulk of the stellar mass is likely located. Indeed, the dashed circle in Fig. \ref{fig:maps} corresponds to the effective radius measured by the Near-Infrared Camera (NIRCam) on board of JWST for the stellar distributions in $z\gtrsim6$ quasars \citep[e.g.][]{Ding23, Stone23, Stone24,Yue24}, powered by black holes with similar $L_{\rm Bol}$ and $M_{\rm BH}$ to \psoj.

The [CII] emission in \psoj, detected at $\sim50\sigma_{\rm [CII]}$ level, where $\sigma_{\rm [CII]}=0.015$ \jybeamkms, spans an angular region of almost $2\times2$ arcsec$^2$ around the quasar location (Fig. \ref{fig:maps}b) and appears to be clumpier than the continuum emission \citep[e.g.][]{Venemans19, Zanella24}.  We measure a [CII] luminosity \citep[using Eq. 1 in ][]{Solomon05} $L_{\rm [CII]} \simeq 1.1\times10^{10}$ \lsun\ (Table \ref{tab:cii-properties}), that is a factor of about two higher than previous measurements in this source \citep{Decarli18,Venemans20}. By imaging the individual datasets following the approach in Sect. \ref{sect:data}, we verified that this is due to the increased sensitivity to the extended [CII] emission provided by the combined dataset (Fig. \ref{fig:cii-spectrum}c). By fitting the [CII] map with a 2D Gaussian profile, we find a deconvolved FWHM size of ($0.83\pm0.04$) $\times$ ($0.75\pm0.03$) arcsec$^2$, corresponding to about $4.7\times4.2$ kpc$^2$. However a single component Gaussian fit results in bright residuals (at $\sim15\sigma_{\rm [CII]}$ level, Fig. \ref{fig:cii-profile}a) in the central 0.2 arcsec, and negative residuals at larger distance from the nucleus, suggesting two distinct [CII] emitting components. No significant residuals are obtained by fitting a two Gaussians model (Fig. \ref{fig:cii-profile}b): this results in a compact component, with a deconvolved major FWHM size of (0.26$\pm 0.03$) arcsec ($\sim1.5$ kpc), consistent with the continuum size, plus an extended component with a FWHM of (1.03 $\pm 0.04$) arcsec, corresponding to about 5.8 kpc.
The size of this extended component is a factor of two-to-three larger than the [CII] sizes typically measured using high-resolution \citep[$\lesssim0.2$ arcsec,] []{Bischetti18,Venemans20,Neeleman23} ALMA observations, while it is among the largest sizes reported using moderate or low-resolution ALMA observations, which are more sensitive to [CII] emission on scales $>>1$ kpc \citep{Decarli18,Fudamoto22, Wang24}. The extended [CII] also reaches significantly further out than the stellar distribution measured by JWST (Fig. \ref{fig:maps}b), suggesting that we are probing [CII] emission arising from the interface region between ISM and CGM.

The [CII] velocity map in \psoj\ (Fig. \ref{fig:maps}c) shows a velocity gradient along the East-West direction, consistent with the ISM rotation identified by \cite{Neeleman21}, although deviations are present in the North and mostly in the South regions at $\sim3$ kpc from the quasar. This may be due to radial gas flows: either  inflows close to the minor rotation axis, with a velocity component along the line of sight of $\sim40$ \kms,  or outflowing gas whose presence is revealed by high-velocity [CII] emission (Sect. \ref{sect:outflow}). The
[C II] velocity dispersion is generally moderate ($\sigma_{\rm v}\lesssim100$ \kms), consistent with the [CII] clumps being formed through violent disk instabilities, similarly to what observed in other high-z galaxies and quasars \citep[e.g][]{Forster-Schreiber18,Inoue16}. An increased $\sigma_{\rm v}\sim150-200$ \kms\  is observed in a biconical region centered on the quasar and extending out to $\sim5$ kpc in the North-East to South-West direction. 
We interpret this increased dispersion as due to the interaction of a [CII] outflow with the ISM and CGM  of the host galaxy (Sect. \ref{sect:outflow}), similarly to what observed in low-redshift Seyferts and quasars,  in which it has been possible to spatially disentangle outflow and disk components \citep{Shimizu19,Bischetti19pds,Zanchettin23}. 
Other processes 
such as galaxy interactions are disfavoured as (i) they cannot reproduce such a symmetric structure with high $\sigma_{\rm v}$ \citep[e.g.][]{Decarli19}, and (ii) we do not see a double peaked continuum or a disturbed [CII] morphology, which suggests that \psoj\ is an isolated galaxy, in agreement with \cite{Venemans20,Neeleman21}. We do not observe a systematic decrease in velocity dispersion with increasing distance from the nucleus.

\section{DISCUSSION}
\subsection{The [CII] outflow in \psoj}\label{sect:outflow}
\begin{figure}[H]
    \centering
    \includegraphics[width=0.9\columnwidth]{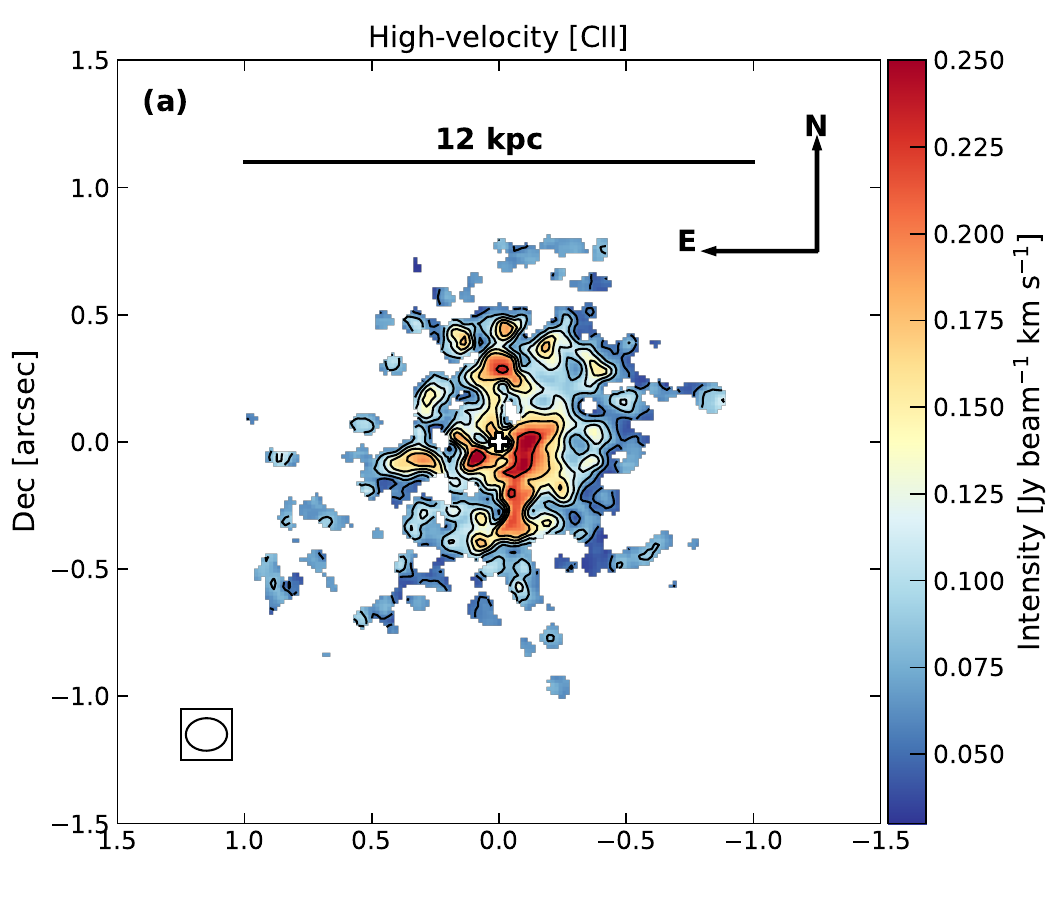}
    \includegraphics[width=0.9\columnwidth]{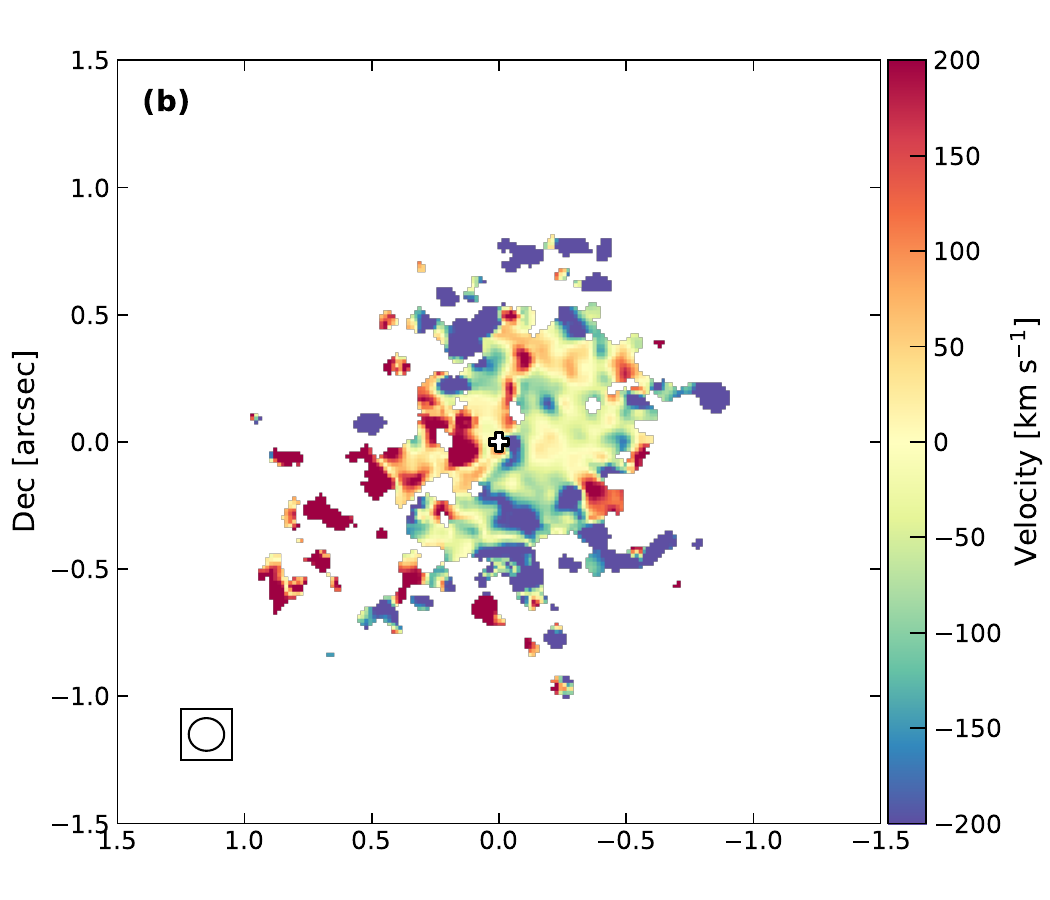}    
    \includegraphics[width=0.9\columnwidth]{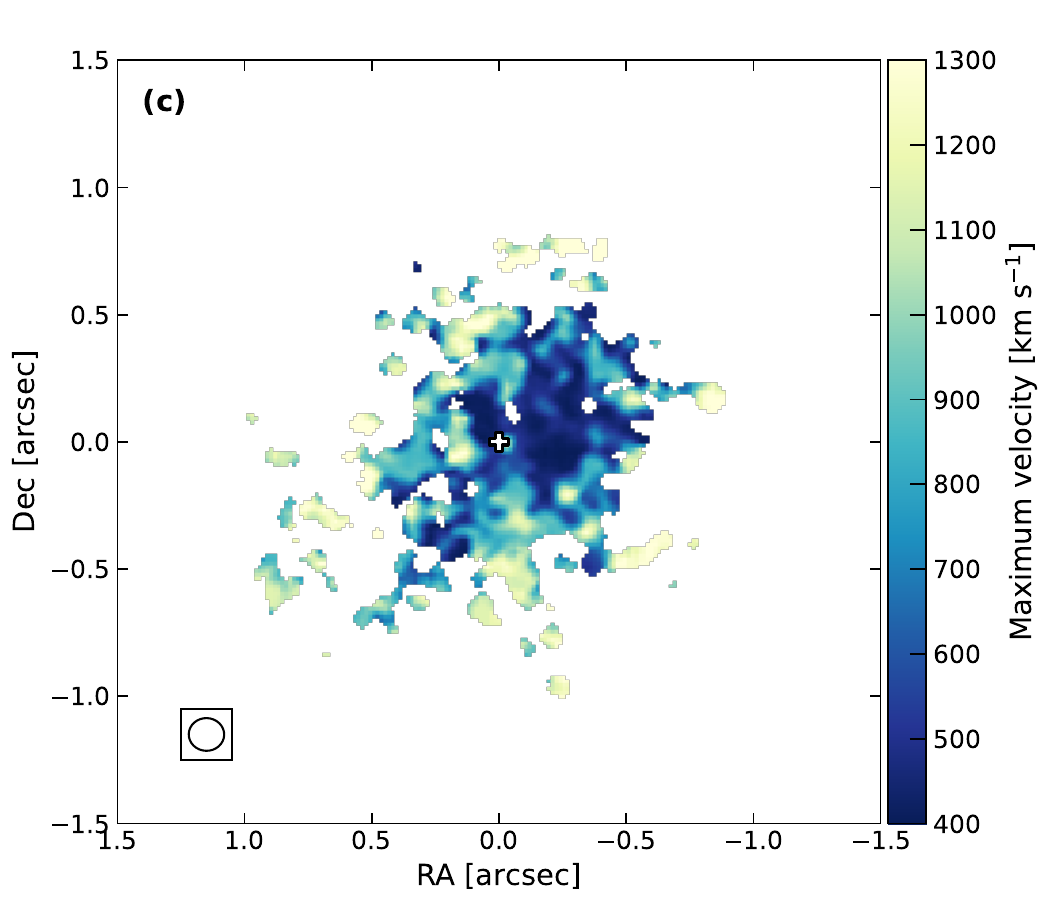}
    \caption{(a): Velocity-integrated intensity map associated with the broad [CII] emission (calculated as in Sect. \ref{sect:data}). Contours correspond to [2,3,4,5,6,8]$\sigma_{\rm [CII]}^{\rm of}$. (b): Velocity map.  (c) Maximum velocity of the [CII] outflow. The cross refers to the quasar location, corresponding to the peak of the ALMA continuum emission (Fig. \ref{fig:maps}a). The white ellipse shows the ALMA beam.}
    \label{fig:outflow-maps}
\end{figure}

Fig. \ref{fig:outflow-maps} shows the moment 0$^{\rm th}$, moment 1$^{\rm st}$ and 2$^{\rm nd}$ maps associated with the broad ($FWHM>500$ \kms) [CII] emission detected (at $\sim8\sigma_{\rm [CII]}^{\rm of}$ level, where $\sigma_{\rm [CII]}^{\rm of}= 0.028$ \jybeamkms\ is the rms noise of the moment 0$^{\rm th}$ map) in the host galaxy of \psoj\ (Sect. \ref{sect:data}). We find this emission to be very clumpy (Fig. \ref{fig:outflow-maps}a) and distributed along the edges of the double cone in which the velocity dispersion of the total [CII] profile is increased (Fig. \ref{fig:maps}d), out to a distance of $\sim5$ kpc from the quasar. 
The velocity shift of this broad component is relatively little ( $-200<v_{\rm mom1}^{\rm broad}<200$ \kms), and the emission is mostly redshifted (blueshifted) in the North-East (South-West) cone. We calculate the maximum velocity as $v_{\rm max}=|v_{\rm mom1}^{\rm broad}|+2\sigma_v^{\rm broad}$ \citep[e.g.][]{Bischetti17,Fiore17}, where $\sigma_v^{\rm broad}$ is the velocity dispersion of the broad [CII] component. We find that 46\% of the gas in the broad component has $v_{\rm max}>750$ \kms, and 18\% reaches $v_{\rm max}$ of $\sim1000-1200$ \kms\ (Fig. \ref{fig:outflow-maps}c). Such velocity is among the highest values measured in the [CII] profiles of local AGN and luminous quasars up to $z\gtrsim6$ \citep{Janssen16,Bischetti19,Izumi21,Izumi21a,Cicone14}, and indicates that the broad [CII] emission cannot be ascribed to bounded motion in the quasar host galaxy, but is instead associated with outflowing gas. 
We note that the highest-velocity gas is not detected in the \cii\ spectrum extracted from the high-resolution dataset alone (Fig. \ref{fig:cii-spectrum}c).

By performing a dynamical modelling of the [CII] kinematics in \psoj, \cite{Neeleman21} reported a dynamical mass $M_{\rm dyn}\sim1.3\times10^{11}$ \msun\ in the inner 4 kpc around the quasar, which corresponds to an escape velocity of about 700 \kms. This implies that a significant fraction of the high-velocity [CII] emission {may be} able to escape the potential well of the host galaxy and reach CGM scales if the outflow propagates ballistically. If the outflow is continuously accelerated, even the lower-velocity gas may escape. At the same time, exceeding the local escape velocity may not be a sufficient condition for reaching the CGM if the outflow entrains a large mass, as it may be expected for the portion of the outflow cone which propagates into the disc plane (Fig. \ref{fig:maps}d).

From the moment 0$^{\rm th}$ map of the high-velocity [CII] we calculate the outflowing atomic gas mass ${M}_{\rm of}\sim5.4\times10^{9}$ \msun\ in the assumption that most of [CII] is excited by photodissociation regions, according to Eq. (1) in \citep{Hailey-Dunsheath10}, for a gas temperature $T=200$K and a density significantly higher than [CII] critical density \citep{Maiolino12,Lagache18,Bischetti19}. This value corresponds to about 22\% of the total atomic mass in \psoj.
We calculate the mass-outflow rate at each radius $r_{\rm of}$ by applying the relation for a conical wind $\dot{M}_{\rm of}=\Omega\frac{M_{\rm of}v_{\rm max}}{r_{\rm of}}f$ \citep[e.g.][]{Bischetti19pds,Bischetti17, Cicone15}, where $\Omega$ is the fractional solid angle spanned by the outflow bicone, and $f\simeq1$ for a density profile scaling as $r_{\rm of}^{-2}$, while $f\simeq3$ for a constant density profile \citep[e.g.][]{Veilleux17,Fiore17}. In our calculation, we calculate $r_{\rm of}$ for each pixel of the outflow map outside the central beam, while we consider $r_{\rm of}$ equal to the beam radius in the central region. We consider $\Omega\sim1/2$ (corresponding to a bicone opening angle of about 120 deg, Fig. \ref{fig:maps}d) and $f\simeq1$. The resulting total atomic mass outflow rate is $\dot{M}_{\rm of}=930^{+330}_{-290}$ \msun\ yr$^{-1}$, where the uncertainty is dominated by that on the bicone opening angle (about $\pm20$ deg). 
This value is among the highest outflow rates reported in $z\gtrsim6$ quasars \citep[e.g.][and references therein]{Tripodi24}. However, it is not much higher than the star-formation rate SFR $\sim650-890$ \msun\ yr$^{-1}$ measured in the host galaxy of \psoj\ based-on multi-frequency ALMA data sampling, and assuming a quasar contribution of 50\% to dust heating \citep{Duras17, Decarli23, Tripodi24sfr}. This implies that a starburst contribution to the outflow acceleration in \psoj\ cannot be a priori excluded \citep[e.g.][]{Fluetsch19, Bischetti19pds}. However, we calculate an outflow kinetic power $\dot{E}_{\rm of}=\frac{1}{2}\dot{M}_{\rm of}v_{\rm max}^2\sim8.3\times10^{43}$ \ergs, corresponding to about 0.5\% of \lbol. This value implies that only a small coupling of the quasar radiation with the host-galaxy medium is sufficient to drive the observed [CII] outflow \citep{Costa14, Richings18}. 

A direct comparison between the [CII] outflow and that detected in OH by \cite{Butler23} is complicated by the uncertain size of the OH outflow, assumed to be $\sim1.2$ kpc based-on the millimeter continuum. \cite{Butler23} also assumes a covering 

\noindent factor of unity, providing an upper limit on the molecular mass outflow rate.
Molecular outflows detected in emission, such as those traced by CO, are typically compact, whereas [CII] outflows can reach several kpc \citep[e.g.][]{Cicone14,Fiore17}, as observed in \psoj. 
As the dynamical timescale inferred from the OH and [CII] outflow is similar (a few Myr), the two phases may trace a same outflow driven by the same black hole accretion or starburst event. In this scenario, the lower OH velocities \citep[500-800 \kms,][]{Butler23} may be interpreted as denser molecular clumps being embedded in a more diffuse, faster neutral outflow \citep[e.g.][]{Richings18}. Accordingly, the total outflow rate could 
reach $\sim2000$ M$_\odot$ yr$^{-1}$.

The coexistence of powerful winds and high SFR has been often observed in the ISM of high-z quasar host galaxies \citep{Feruglio17,Bischetti21,Lamperti21,Vayner24} and expected by cosmological simulations of early galaxy-evolution \citep[e.g.][]{Valentini21}, which is at odds with the expectations for  an ejective feedback mode efficiently removing gas before it
can fuel star formation. Instead, recent feedback theories have shifted toward a  delayed feedback mode, primarily associated with processes occurring on CGM scale. The underlying idea is that the
physical and kinematic conditions of the CGM may change during the feedback process, and these alterations may be the primary mechanism influencing further gas accretion and, consequently, reducing star
formation \citep{Barai18,Costa22}. 
Such a scenario may be reasonably applied to \psoj, as the [CII] wind is able to reach and propagate its energy and momentum to the CGM gas on a short timescale of a few million years, given the observed $v_{\rm max}$.

\begin{figure}[thb]
    \centering
    \includegraphics[width=0.48\columnwidth]{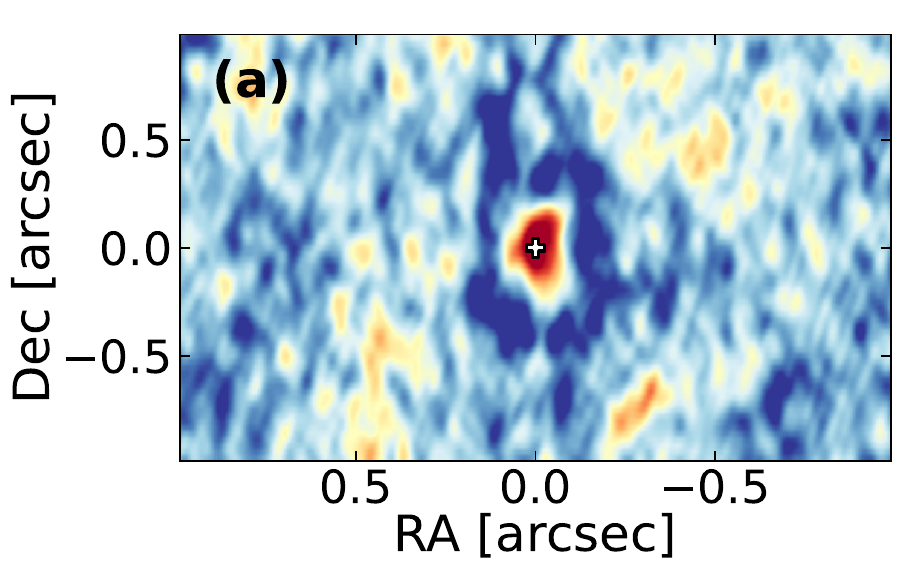}
    \includegraphics[width=0.48\columnwidth]{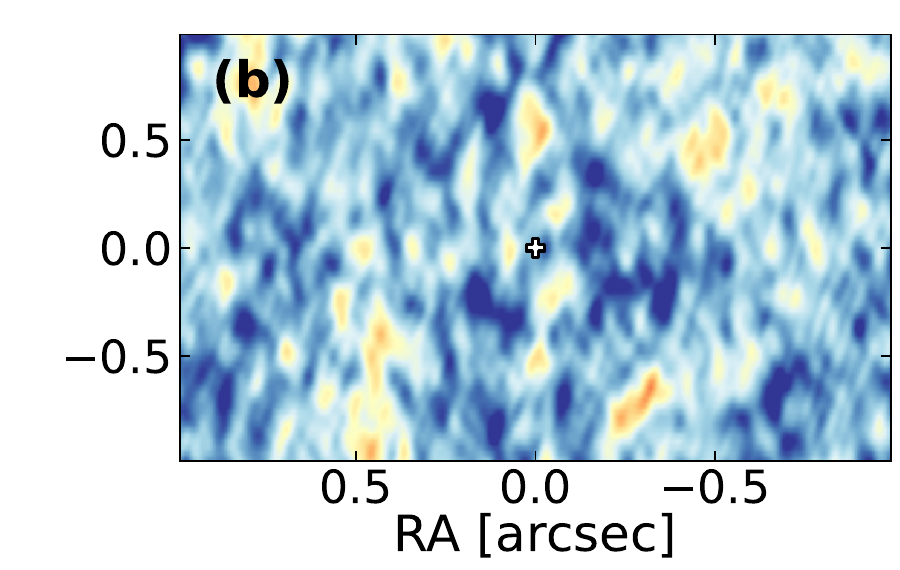}
    \includegraphics[width=1\columnwidth]{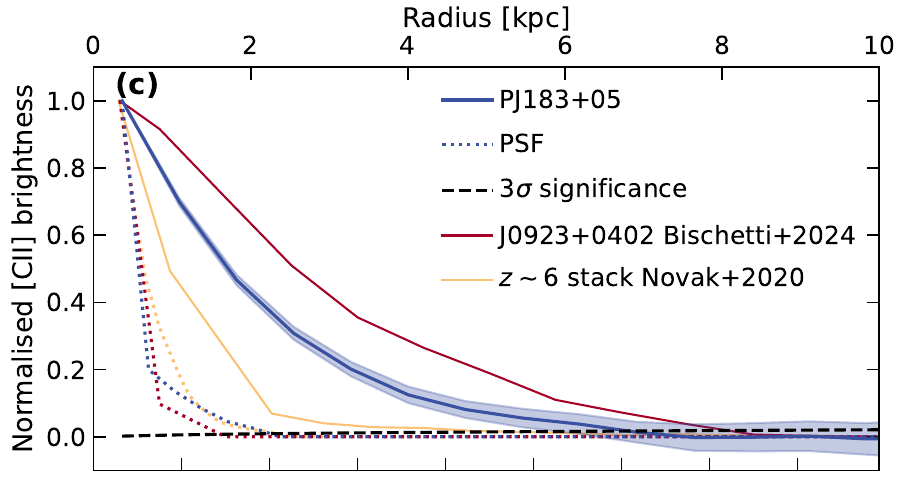}
    \includegraphics[width=1\columnwidth]{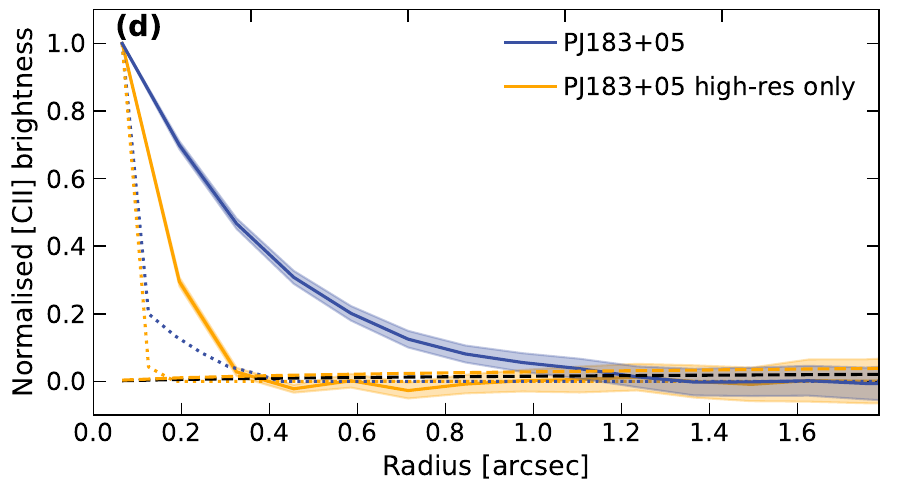}
    \caption{Residual map obtained by fitting a single (a) or a double (b) 2D Gaussian profile to the [CII] moment 0$^{\rm th}$ map (Sect. \ref{sect:results}). (c): Brightness profile of the [CII] emission in \psoj\ (blue solid curve), normalised to the central peak value, and associated 68\% confidence level uncertainty (shaded area). The brightness profile associated with the ALMA point spread function (PSF) is shown by the dotted blue curve, while the dashed curve shows the profile associated with the $3\sigma$ level. We also show the profile for another $z\sim6.6$ quasar (in a merger, red curve) showing extended [CII] emission from \cite{Bischetti24} and the stacked profile of $z\sim6$ quasar hosts by \cite{Novak20}, shown by the yellow curve. (d) [CII] brightness profile of \psoj\ as in panel (c), compared to that extracted from the high-resolution dataset alone (Sect. \ref{sect:data}).}
    \label{fig:cii-profile}
\end{figure}

\subsection{Bright CGM emission and feedback}\label{sect:halo}

The ALMA observations of \psoj\ presented in this work reveal the presence of a bright [CII] emission associated with the host galaxy ISM and CGM. [CII] is mostly a tracer of neutral atomic gas in photon-dominated regions (PDRs) around young stars but it can be also emitted from the partly ionised medium \citep{Lagache18, Casavecchia24}. However, the larger [CII] extent with respect to the millimetre continuum (Fig. \ref{fig:maps}a,b) suggests that a significant fraction of the extended [CII] emission may not arise from PDRs but may rather trace diffuse and ionised gas in the CGM.

This is supported by the fact that about 50\% of the [CII] emission arises beyond a radius of 0.5 arcsec ($\simeq2.8$ kpc), as it can be seen from Fig. \ref{fig:maps}b and the [CII] brightness profile of \psoj\ shown in Fig. \ref{fig:cii-profile}c. The profile shows the average [CII] emission in circular annuli of $0.13"$ radius, calculated following the method described in \cite{Tripodi22, Bischetti24}. 

[CII] emission on scales beyond a few kpc has been previously detected by stacking samples of star forming galaxies at $z\gtrsim5$, and in a few individual targets \citep{Fujimoto19,Fujimoto20, Ginolfi20, Herrera-Camus21, Akins22, Lambert23}. 
We find that the size of the [CII] emission around \psoj\ is about twice more extended than the stacked [CII] profile of $z\sim6$ quasars by \cite{Novak20}. However, we note that the latter profile was based on high-resolution only observations, which are not optimized to detect possible diffuse [CII] emission. Indeed, by extracting the [CII] brightness profile of \psoj\ from the high-resolution dataset alone (Sect. \ref{sect:data}), we find that about 55\% of the total [CII] emission and 100\% of the emission beyond a 2 kpc radius is not recovered (Fig. \ref{fig:cii-profile}d).
This is in agreement with ALMA simulated observations of $z>6$ galaxies by \cite{Carniani20}, showing that a major fraction (about 50\%) of the flux of the extended [CII] emission is not recovered when employing high-resolution observations only.  By combining ALMA observations acquired with different antenna configurations \cite{Bischetti24} reported the discovery of a bright [CII] emission in the CGM of $z\sim6.6$ quasar J0923$+$0402.
The size of the [CII] emission in \psoj\ is smaller than that of J0923$+$0402, although the latter is in a merging system, which might contribute to increase the region of [CII] emitting gas \citep{Ginolfi20b, Lambert23}.

Recent works have suggested a link between feedback and CGM halos around $z\sim6$ star forming galaxies and quasars. Given the presence of a large-scale [CII] outflow in \psoj, such a scenario might explain its extended [CII] emission. Feedback via high-velocity outflows can displace gas beyond a few kpc \citep{Costa19,Vito22} and, at the same time, outflows can significantly contribute to gas heating via shocks \citep{Appleton13,Fujimoto20, Pizzati23}. 
The multi-phase structure of the CGM gas depends on the fraction of photons reaching the CGM scales, which in turn depends on black-hole feedback clearing out the lines of sight \citep{Costa22}. Along the lines of sight with high escape fraction, the halo would be mostly ionized \citep[e.g.,][]{Barai18, Obreja24}, consistently with the Ly$\alpha$ halos frequently detected around high-z quasars \citep{Borisova16,Farina19}. If the global escape fraction remains relatively low \citep{Stern21}, extended and bright [CII] halos are expected in massive halos such as those of $z\gtrsim6$ quasars \citep{Costa23,Pizzati23}.

\section{CONCLUSIONS}
This study has demonstrated that by combining ALMA datasets acquired with different antenna configurations, extended [CII] emission and can be detected beyond ISM scale around the host galaxies of $z\gtrsim6$ quasars. At the same time, mapping the cold gas emission with high-resolution allows to probe the gas kinematics on a broad range of scales ($\sim500$ pc to several kpc) and to investigate the presence of inflows and outflows and assess their impact on the large-scale gas reservoir. A major fraction of the total \cii\ emission is missed when relying on the high-resolution dataset alone. This highlights a well-known limitation of interferometric observations in the low-redshift Universe, where galaxies often exhibit large angular sizes compared to the ALMA beam. In such cases, combining multiple array configurations is a commonly adopted strategy \citep[e.g.][]{Garcia-Burillo21, RamosAlmeida22}. However, this approach has been rarely applied at high redshift \citep[e.g.][]{Carniani20}. Our findings suggest that this limitation may underlie the conflicting results reported in the literature regarding the extent of cold gas reservoirs and the detection of \cii\ outflows in high-redshift galaxies \citep[e.g.][]{Cicone15, Bischetti19, Novak20, Meyer22}.

Several quasars and high-z galaxies have already been observed with multiple antenna configurations, which suggests that significant information on the CGM of massive early galaxies lies unexplored in the ALMA archive \citep[e.g. see also][]{Bischetti24}. However, the archival observations contains heterogenous observations in terms of angular resolution, sensitivity, and frequency coverage, which are not suited for a systematic study of CGM properties. This implies that complementary dedicated ALMA observations are needed. 

To build a three-dimensional picture of the multi-phase CGM gas, ALMA information can be combined with that provided by MUSE and JWST for the warm ionized gas. In the case of \psoj, the high redshift implies that Ly$\alpha$ transition lies close to the edge of the MUSE spectral coverage, 
making a detection of diffuse Ly$\alpha$ more difficult. Indeed, a Ly$\alpha$ CGM halo was not detected in this quasar in a 0.8 hours exposure \citep{Farina19}. However, \psoj\ is being followed up with deeper observations as part of the MUSE program 112.262L.002 (PI E. Farina). In addition, upcoming Cycle 3 JWST observations with NIRSpec IFU (GO program 4912, PI S. Carniani) will allow us to map the morphology, metal enrichment and kinematics in the host-galaxy and up to the CGM of \psoj, using rest-frame optical tracers such as H$\alpha\ \lambda6564$ \AA, [OIII] $\lambda5008$ \AA, and other metal  lines \citep{Marshall23,Liu24,Decarli24}. 
Such an approach is key to build a detailed picture of how baryons cycle between galaxies and their CGM, and improve the current understanding of the evolution of the first massive galaxies.



\vspace{0.5cm}


\begin{acknowledgments}
This paper makes use of the following ALMA data: ADS/JAO.ALMA\#2019.1.01633.S (P.I. M. Neeleman), ADS/JAO.ALMA\#2016.1.00544.S (P.I. E. Banados), ADS/JAO.ALMA\#2015.1.01115.S (P.I. F. Walter), and ADS/JAO.ALMA\#2021.1.01082.S (P.I. S. Bosman). ALMA is a partnership of ESO (representing its member states), NSF (USA), and NINS (Japan), together with NRC (Canada), MOST and ASIAA (Taiwan), and KASI (Republic of Korea), in cooperation with the Republic of Chile. The Joint ALMA Observatory is operated by ESO, AUI/NRAO and NAOJ. This work is based in part on observations made with the NASA/ESA/CSA James Webb Space Telescope. The project leading to this publication has received support from ORP, that is funded by the European Union’s Horizon 2020 research and innovation programme under grant agreement No 101004719 [ORP]. M.B. acknowledges support from INAF project 1.05.12.04.01 - MINI-GRANTS di RSN1 "Mini-feedback" and from UniTs under FVG LR 2/2011 project D55-microgrants23 "Hyper-gal". M.B., C.F., F.F., and F.S. acknowledge support from INAF PRIN 2022 2022TKPB2P "BIG-z", INAF Bando Ricerca Fondamentale 2023 Data grant "ARCHIE", and M4C2 Missione 4 “Istruzione e Ricerca” - Componente C2 Investimento 1.1 Fondo per il Programma Nazionale di Ricerca e Progetti di Rilevante Interesse Nazionale (PRIN) prog. PRIN 2022 PNRR P2022ZLW4T, "Next-generation computing and data technologies to probe the cosmic metal content". F.S. acknowledges financial support from Ricerca Fondamentale INAF 2024 under project "ECHOS" MINI-GRANTS RSN1. 

\end{acknowledgments}

%

\vspace{5mm}
\facilities{ALMA}.


\software{astropy \citep{Astropy13,Astropy18,Astropy22},  
          }



\clearpage

\bibliography{biblio}{}
\bibliographystyle{aasjournal}



\end{CJK}
\end{document}